\documentstyle[sprocl]{article}
\def\be{\begin{equation}}
\def\ee{\end{equation}}
\def\bea{\begin{eqnarray}}
\def\eea{\end{eqnarray}}

\begin{document}
\title{INFLATION AND THE THEORY OF COSMOLOGICAL PERTURBATIONS\footnote{Invited lectures at the 1997 ICTP Trieste Summer School on Particle Physics and Cosmology, to be published in the proceedings}}
\author{ ROBERT H. BRANDENBERGER }
\address{Brown University Physics Department\\
Providence, RI 02912, USA}
%
%

\maketitle
\abstracts{The hypothesis that the Universe underwent a period of exponential expansion at very early times has become the most popular theory of the early Universe. Not only does it solve some of the problems of standard big bang cosmology, but it also provides a causal theory for the origin of primordial density fluctuations which may explain the observed density inhomogeneities and cosmic microwave fluctuations in the Universe. In these lectures, a review of the basic principles of inflationary cosmology is given, focusing on the theory of the origin and growth of cosmological perturbations. The lectures also focus on some recent progress in inflationary cosmology (in particular on an improved theory of reheating), and on several problems with the present inflationary Universe models. A couple of possible approaches to resolve these issues are suggested.}

\medskip
\noindent{Brown preprint BROWN-HET-1098}
\medskip

\section{The Inflationary Universe Scenario}

\subsection{Standard Cosmology: Pillars, Tests and Problems}
 
The standard big bang cosmology rests on three theoretical pillars: the
cosmological principle, Einstein's general theory of relativity and a perfect
fluid description of matter.

The cosmological principle states that on large distance scales the
Universe is homogeneous and isotropic. This implies that space-time is described by a Robertson-Walker metric (see e.g. \cite{Wein}).
The dynamics of an expanding Universe  is then given by the Friedmann-Robertston-Walker (FRW) equations
\be
\left( {\dot a \over a} \right)^2 - {k\over a^2} = {8 \pi G\over 3 } \rho
\ee
\be
{\ddot a\over a} = - {4 \pi G\over 3} \, (\rho + 3 p) \, ,
\ee
where $\rho$ is the energy density and $p$ is the pressure.  
The third key assumption of standard cosmology is that matter is described by
an ideal gas with an equation of state
\be
p = w \rho \, . 
\ee
 
The three classic observational pillars of standard cosmology are Hubble's law, the existence and black body nature of the nearly isotropic CMB, and the abundances of light elements (nucleosynthesis). These successes are discussed in detail in many textbooks on cosmology (see e.g. \cite{Wein}), and will therefore not be reviewed here.

Since the energy density in radiation redshifts faster than the matter energy density, it follows by working backwards in time from the present data that although the energy density of the Universe is now mostly in cold matter, it was initially dominated by radiation. The transition occurred at a time denoted by $t_{eq}$, the ``time of equal matter and radiation". As will be discussed in Section 2, $t_{eq}$ is the time when perturbations can start to grow by gravitational clustering. The second important time is $t_{rec}$, the ``time of recombination" when photons fell out of equilibrium. For the usual values of the cosmological parameters, $t_{eq} < t_{rec}$. 

Standard Big Bang cosmology is faced with several important problems.  Only one
of these,  the age problem, is a potential conflict with observations.  The
others which I will focus on here -- the homogeneity, flatness and formation of
structure problems (see e.g. \cite{Guth}) -- are questions which have no answer
within the theory and are therefore the main motivation for the inflationary Universe scenario. Because of lack of space, the discussion of the problems of standard cosmology and of their solution in the context of inflation is very brief. For an extended discussion (with useful figures) see e.g. \cite{TASI}.

The horizon problem is the fact that in standard cosmology the comoving
region $\ell_p (t_{rec})$ over which the CMB is observed to be homogeneous (to
better  than one part in $10^4$) is much larger than the comoving forward light
cone $\ell_f (t_{rec})$ at $t_{rec}$, which is the maximal distance over which
microphysical forces could have caused the homogeneity:
\be
\ell_p (t_{rec}) = \int\limits^{t_0}_{t_{rec}} dt \, a^{-1} (t) \simeq 3 \, t_0
\left(1 - \left({t_{rec}\over t_0} \right)^{1/3} \right) 
\ee
\be
\ell_f (t_{rec}) = \int\limits^{t_{rec}}_0 dt \, a^{-1} (t) \simeq 3 \, t^{2/3}_0
\, t^{1/3}_{rec} \, . 
\ee
{}From the above equations it is obvious that $\ell_p (t_{rec}) \gg \ell_f
(t_{rec})$.  Hence, standard cosmology cannot explain the observed isotropy of
the CMB.

In standard cosmology and in an expanding Universe, $\Omega = 1$ is an unstable
fixed point. From the FRW equations it follows that 
\be \label{omega3}
\Omega - 1 \, = \, {\rho - \rho_c\over \rho_c} = {3\over{8 \pi G}} \, {\varepsilon T^2\over
\rho_c} \sim T^{-2} \, . 
\ee 
where $\rho_c$ is the energy density of a spatially flat Universe and
\be
\varepsilon = {k\over{(aT)^2}} \,  
\ee
($k = \pm 1, 0$ is the spatial curvature constant). In standard cosmology, $\varepsilon$ is constant.
Thus, as the temperature decreases, $\Omega - 1$ increases.  In fact, in order
to explain the present small value of $\Omega - 1 \sim {\cal O} (1)$, the
initial energy density had to be extremely close to critical density. This is the flatness problem of standard cosmology.

The third of the classic problems of standard cosmological model is the
``formation of structure problem."  Observations indicate that galaxies and
even clusters of galaxies have nonrandom correlations on scales larger than 50
Mpc (see e.g. \cite{CFA,LCRS}).  This scale is comparable to the comoving horizon at
$t_{eq}$.  Thus, if the initial density perturbations were produced much before
$t_{eq}$, the correlations cannot be explained by a causal mechanism.  Gravity
alone is, in general, too weak to build up correlations on the scale of
clusters after $t_{eq}$.
Hence, the two questions of what generates the primordial density perturbations
and what causes the observed correlations, do not have an answer in the context
of standard cosmology.   
 
\subsection{Inflation as a Solution}

The idea of inflation \cite{Guth} is very simple (for some early reviews of inflation see e.g. \cite{Linde,GuthBlau,Olive,RB85}).  We assume there is a time
interval beginning at $t_i$ and ending at $t_R$ (the ``reheating time") during
which the Universe is exponentially expanding, i.e.,
\be
a (t) \sim e^{Ht}, \>\>\>\>\> t \epsilon \, [ t_i , \, t_R] 
\ee
with constant Hubble expansion parameter $H$.  Such a period is called  ``de
Sitter" or ``inflationary."  The success of Big Bang nucleosynthesis requires that reheating occur before nucleosynthesis. 

There are several phases of an inflationary Universe.  Before the
onset of inflation there are no constraints on the state of the Universe.  In
some models a classical space-time emerges immediately in an inflationary
state, in others there is an initial radiation dominated FRW period. After the inflationary phase which ends at $t_R$, the Universe is very hot and
dense, and the subsequent evolution is as in standard cosmology.  During the
inflationary phase, the number density of any particles initially in thermal
equilibrium at $t = t_i$ decays exponentially.  Hence, the matter temperature
$T_m (t)$ also decays exponentially.  At $t = t_R$, all of the energy which is
responsible for inflation (see later) is released as thermal energy.  This is a
nonadiabatic process during which the entropy increases by a large factor.   

It is quite evident how a period of inflation can solve the homogeneity
problem.  $\Delta t = t_R - t_i$  is the period of inflation.  During
inflation, the forward light cone increases exponentially compared to a model
without inflation, whereas the past light cone is not affected for $t \geq
t_R$.  Hence, provided $\Delta t$ is sufficiently large, $\ell_f (t_R)$ will be
greater than $\ell_p (t_R)$. 

Inflation also can solve the flatness problem \cite{Kazanas,Guth}. The key point is
that the entropy density $s$ is no longer constant.  As will be explained
later, the temperatures at $t_i$ and $t_R$ are essentially equal.  Hence, the
entropy increases during inflation by a factor $\exp (3 H \Delta t)$.  Thus,
$\epsilon$ decreases by a factor of $\exp (-2 H \Delta t)$.  Hence, $(\rho - \rho_c) / \rho$ can be of order 1 both at $t_i$ and at the
present time.  In fact, if inflation occurs at all, then rather generically, the theory predicts
that at the present time $\Omega = 1$ to a high accuracy (now $\Omega < 1$
requires  special initial conditions or rather special models \cite{open}).

Most importantly, inflation provides a mechanism which in a causal way
generates the primordial perturbations required for galaxies, clusters and even
larger objects.  In inflationary Universe models, the Hubble radius
(``apparent" horizon), $H^{-1}$ is much smaller than the particle horizon (the forward light cone).
Provided that the duration of inflation is sufficiently long, then all scales within the present Hubble radius were inside the particle
horizon at the beginning of inflation.  Thus, it is in principle possible to have a casual
generation mechanism for perturbations \cite{Press,Mukh80,Lukash,Sato}.

The generation of perturbations is supposed to be due to a causal microphysical
process.  Such processes can only act coherently on length scales smaller than
the Hubble radius $H^{-1} (t)$. This can be seen, for example, by considering the wave equation for a scalar field and observing that on scales larger than the Hubble radius, the Hubble damping term dominates the dynamics.

As will be discussed in Section 3, the density perturbations produced during
inflation are due to quantum fluctuations in the matter and gravitational
fields \cite{Mukh80,Lukash}.  The amplitude of these inhomogeneities corresponds to the Hawking temperature $T_H  \sim H$ of the de Sitter phase. This implies that at all times
$t$ during inflation, perturbations with a fixed physical wavelength $\sim
H^{-1}$ will be produced. Subsequently, the length of the waves is stretched
with the expansion of space, and soon becomes larger than the Hubble radius.
The phases of the inhomogeneities are random.  Thus, the inflationary Universe
scenario predicts perturbations on all scales ranging from the comoving Hubble
radius at the beginning of inflation to the corresponding quantity at the time
of reheating.  In particular, provided that inflation lasts sufficiently long, perturbations on scales of galaxies and beyond will be generated.  

\subsection{Models of Inflation}

Obviously, the key question is how to obtain inflation. From the FRW equations, it follows that in order to get exponential increase of the scale factor, the equation of state of matter must be
\be \label{infleos}
p = - \rho  
\ee
This is where the connection with particle physics comes in. The energy density and pressure of a scalar quantum field $\varphi$ are given by
\begin{eqnarray}
\rho (\varphi) & = & {1\over 2} \, \dot \varphi^2 + {1\over 2} \,
(\nabla \varphi)^2 + V (\varphi) \label{eos1} \\
p (\varphi) & = & {1\over 2} \dot \varphi^2 - {1\over 6} (\nabla \varphi)^2 - V
(\varphi) \, . \label{eos2}
\end{eqnarray}
Thus, provided that at some initial time $t_i$
\be \label{incond}
\dot \varphi (\underline{x}, \, t_i) = \nabla \varphi (\underline{x}_i \, t_i)
= 0 
\ee
and
\be
V (\varphi (\underline{x}_i, \, t_i) )  > 0 \, , 
\ee
the equation of state of matter will be (\ref{infleos}).
 
The next question is how to realize the required initial conditions (\ref{incond}) and
to maintain the key constraints
\be
\dot \varphi^2 \ll V (\varphi) \> , \> (\nabla \varphi)^2 \ll V (\varphi)
\ee
for sufficiently long. Various ways of realizing these conditions were put forward, and they gave rise to different models of inflation. I will focus on ``old inflation," ``new inflation"" and ``chaotic
inflation."   

\smallskip
\noindent{\it Old Inflation}
 
The old inflationary Universe model \cite{Guth,GuthTye}  is based on a scalar field
theory which undergoes a first order phase transition.  As a toy
model, consider a scalar field theory with a  potential $V (\varphi)$
which has a metastable ``false" vacuum at $\varphi = 0$, whereas the lowest energy state (the ``true" vacuum) is $\varphi = a$. Finite temperature effects \cite{finiteT}  lead to extra terms in the finite temperature effective potential which are proportional to $\varphi^2 T^2$.  Thus, at high temperatures, the energetically preferred state is the false vacuum state. Note that this is only true if $\varphi$ is in thermal equilibrium with the other fields in the system.
 
For fairly general initial conditions, $\varphi (x)$ is trapped in the
metastable state $\varphi = 0$ as the Universe cools below the
critical temperature $T_c$.  As the Universe expands further, all
contributions to the energy-momentum tensor $T_{\mu \nu}$ except for
the contribution
\be
T_{\mu \nu} \sim V(\varphi) g_{\mu \nu} 
\ee
redshift.  Hence, if $V(a) = 0$, then the equation of state in the false vacuum approaches $p = - \rho$, and
inflation sets in. After a period $\Gamma^{-1}$, where $\Gamma$ is the tunnelling rate, bubbles of $\varphi = a$ begin to nucleate \cite{decay} in a sea of false
vacuum $\varphi = 0$. Inflation lasts until the false vacuum decays.
During inflation, the Hubble constant is given by
\be
H^2 = {8 \pi G\over 3} \, V (0) \, . 
\ee
Note that the condition $V(a) = 0$, which looks rather unnatural, is required to
avoid a large cosmological constant today (none of the present inflationary Universe
models manages to circumvent or solve the cosmological constant problem).
 
It was immediately realized that old inflation has a serious ``graceful exit"
problem \cite{Guth,GuthWein}. The bubbles nucleate after inflation with radius $r \ll
2t_R$ and would today be much smaller than our apparent horizon.  Thus, unless
bubbles percolate, the model predicts extremely large inhomogeneities inside
the Hubble radius, in contradiction with the observed isotropy of the
microwave background radiation.
\par
For bubbles to percolate, a sufficiently large number must be produced so that
they collide and homogenize over a scale larger than the present Hubble
radius.  However, with exponential expansion, the volume between bubbles
expands
exponentially whereas the volume inside bubbles expands only with a low power.
This prevents percolation.

\smallskip
\noindent{\it New Inflation}
 
Because of the graceful exit problem, old inflation never was considered to be
a viable cosmological model.  However, soon after the seminal paper by
Guth, Linde \cite{Linde82}  and independently Albrecht and Steinhardt \cite{AS82} put forwards a modified scenario, the New Inflationary Universe.

The starting point is a scalar field theory with a double well potential which
undergoes a second order phase transition.  $V(\varphi)$ is
symmetric and $\varphi = 0$ is a local maximum of the zero temperature
potential.  Once again, it was argued that finite temperature effects confine
$\varphi(x)$ to values near $\varphi = 0$ at temperatures $T \geq
T_c$.  For $T < T_c$, thermal fluctuations trigger the instability of $\varphi
(x) = 0$ and $\varphi (x)$ evolves towards either of the global minima at $\varphi = \pm \sigma$ by the classical equation of motion
\be \label{eom}
\ddot \varphi + 3 H \dot \varphi - a^{-2} \bigtriangledown^2 \varphi = -
V^\prime (\varphi)\, .  
\ee

Within a fluctuation region, 
$\varphi(x)$ will be homogeneous. In such a region, we can  neglect the spatial gradient terms in Eq. (\ref{eom}).  Then, from (\ref{eos1}) and (\ref{eos2}) we can read off the
induced equation of state.  The condition for inflation is
\be
\dot \varphi^2 \ll V (\varphi)\, , 
\ee
i.e.~ slow rolling.
Often, the  ``slow rolling" approximation is made to find solutions of
(\ref{eom}).
This consists of dropping the $\ddot \varphi$ term.   
 
There is no graceful exit problem in the new inflationary Universe.  Since the
fluctuation domains are established before the onset of inflation,
any boundary walls will be inflated outside the present Hubble radius.

Let us, for the moment, return to the general features of the new inflationary
Universe scenario.  At the time $t_c$ of the phase transition, $\varphi (t)$
will start to move from near $\varphi = 0$ towards either $\pm \sigma$ as
described by the classical equation of motion, i.e.~ (\ref{eom}).  At or soon after
$t_c$, the energy-momentum tensor of the Universe will start to be dominated
by $V(\varphi)$, and inflation will commence.  $t_i$ shall denote the time of
the onset of inflation.  Eventually, $\phi (t)$ will reach large values for which
nonlinear effects become important.  The time at which this occurs is $t_B$.
For $t > t_B \, , \, \varphi (t)$ rapidly accelerates, reaches $\pm \sigma$,
overshoots and starts oscillating about the global minimum of $V (\varphi)$.
The amplitude of this oscillation is damped by the expansion of the Universe
and (predominantly) by the coupling of $\varphi$ to other fields.  At time
$t_R$,
the energy in $\varphi$ drops below the energy of the thermal bath of
particles produced during the period of oscillation.
 
In order to obtain inflation, the potential $V(\varphi)$ must be very flat near the false vacuum at $\varphi = 0$. This can only be the case if all of the coupling constants appearing in the potential are small. However, this implies that $\varphi$ cannot be in thermal equilibrium at early times, which would be required to localize $\varphi$ in the false vacuum. In the absence of thermal equilibrium, the initial conditions for $\varphi$ are only constrained by requiring that the total energy density in $\varphi$ not exceed the total energy density of the Universe. Most of the phase space of these initial conditions lies at values of $| \varphi | >> \sigma$. This leads to the ``chaotic" inflation scenario \cite{Linde83}.

\smallskip
\noindent{\it Chaotic Inflation}
 
Consider a region in space where at the initial time $\varphi (x)$
is very large, homogeneous and static.  In this case, the energy-momentum tensor will be
immediately dominated by the large potential energy term and induce an
equation of state $p \simeq - \rho$ which leads to inflation.  Due to the
large Hubble damping term in the scalar field equation of motion, $\varphi
(x)$ will only roll very slowly towards $\varphi = 0$.  The
kinetic energy contribution to $T_{\mu \nu}$ will remain small, the spatial
gradient contribution will be exponentially suppressed due to the expansion of
the Universe, and thus inflation persists. Note that in contrast to old and new inflation,
no initial thermal bath is required.  Note also that the precise form of
$V(\varphi)$ is irrelevant to the mechanism.  In particular, $V(\varphi)$ need
not be a double well potential.  This is a significant advantage, since for
scalar fields other than Higgs fields used for spontaneous symmetry breaking,
there is no particle physics motivation for assuming a double well potential,
and since the inflaton (the field which gives rise to inflation) cannot be a
conventional Higgs field due to the severe fine tuning constraints.
 
Chaotic inflation is a much more radical departure from standard cosmology than old and new inflation. In the latter, the inflationary phase can be viewed as a short phase of exponential expansion bounded at both ends by phases of radiation domination. In chaotic inflation, a piece of the Universe emerges with an inflationary equation of state immediately after the quantum gravity (or string) epoch.

The chaotic inflationary Universe scenario has been developed in great detail (see e.g. \cite{Linde94} for a recent review). One important addition is the inclusion of stochastic noise \cite{Starob87}  in the equation of motion for $\varphi$ in order to take into account the effects of quantum fluctuations. It can in fact be shown that for sufficiently large values of $|\varphi|$, the stochastic force terms are more important than the classical relaxation force $V^\prime(\varphi)$. There is equal probability for the quantum fluctuations to lead to an increase or decrease of $|\varphi|$. Hence, in a substantial fraction of comoving volume, the field $\varphi$ will climb up the potential. This leads to the conclusion that chaotic inflation is eternal. At all times, a large fraction of the physical space will be inflating. Another consequence of including stochastic terms is that on large scales (much larger than the present Hubble radius), the Universe will look extremely inhomogeneous.

\subsection{Reheating in Inflationary Cosmology}

Recently, there has been a complete change in our understanding of reheating in inflationary cosmology. Reheating is the process of energy transfer between the inflaton and matter fields which takes place at the end of inflation.  

According to either new inflation or chaotic inflation, the dynamics of the inflaton leads first to a transfer of energy from potential energy of the inflaton to kinetic energy. After the period of slow rolling, the inflaton $\varphi$ begins to oscillate about the true minimum of $V(\varphi)$. Quantum mechanically, the state of homogeneous oscillation corresponds to a coherent state. Any coupling of $\varphi$ to other fields (and even self coupling terms of $\varphi$) will lead to a decay of this state. This corresponds to particle production. The produced particles will be relativistic, and thus at the conclusion of the reheating period a radiation dominated Universe will emerge.

The key questions are by what mechanism and how fast the decay of the coherent state takes place. It is important to determine the temperature of the produced particles at the end of the reheating period. The answers are relevant to many important questions regarding the post-inflationary evolution. For example, it is important to know whether the temperature after reheating is high enough to allow GUT baryogenesis and the production of GUT-scale topological defects. In supersymmetric models, the answer determines the predicted abundance of gravitinos and other moduli fields.

Reheating is usually studied using simple scalar field toy models. The one we will adopt here consists of two real scalar fields, the inflaton $\varphi$
with Lagrangian
\be
{\cal L}_o \, = \, {1 \over 2} \partial_\mu \varphi \partial^\mu \varphi - {1 \over 4} \lambda (\varphi^2 - \sigma^2)^2 
\ee
interacting with a massless scalar field $\chi$ representing ordinary matter. The interaction Lagrangian is taken to be
\be
{\cal L}_I \, = \, {1 \over 2} g^2 \varphi^2 \chi^2 \, .
\ee
Self interactions of $\chi$ are neglected. 

By a change of variables
\be
\varphi \, = \, {\tilde \varphi} + \sigma \, ,
\ee
the interaction Lagrangian can be written as
\be \label{intlag}
{\cal L}_I \, = \, g^2 \sigma {\tilde \varphi} \chi^2 + {1 \over 2} g^2 {\tilde \varphi}^2 \chi^2 \, .
\ee
During the phase of coherent oscillations, the field ${\tilde \varphi}$ oscillates with a frequency
\be
\omega \, = \, m_{\varphi} \, = \, \lambda^{1/2} \sigma 
\ee
(neglecting the expansion of the Universe which can be taken into account as in \cite{KLS94,STB95}).

\smallskip
\noindent{\it Elementary Theory of Reheating}
 
According to the elementary theory of reheating (see e.g. \cite{DolLin} and \cite{AFW}), the decay of the inflaton is calculated using first order perturbations theory. According to the Feynman rules, the decay rate $\Gamma_B$ of $\varphi$ (calculated assuming that the cubic coupling term dominates) is
given by
\be
\Gamma_B \, = \, {{g^2 \sigma^2} \over {8 \pi m_{\phi}}} \, .
\ee
The decay leads to a decrease in the amplitude of $\varphi$ (from now on we will drop the tilde sign) which can be approximated by adding an extra damping term $\Gamma_B {\dot \varphi}$ to the equation of motion for $\varphi$.
As long as $H > \Gamma_B$, particle production is negligible. When $H = \Gamma_B$ reheating occurs, i.e. the remaining energy density in $\varphi$ is very quickly transferred to $\chi$ particles.

The temperature $T_R$ at the completion of reheating can be estimated by computing the temperature of radiation corresponding to the value of $H$ at which $H = \Gamma_B$. From the FRW equations it follows that
\be
T_R \, \sim \, (\Gamma_B m_{pl})^{1/2} \, .
\ee
If we now use the ``naturalness" constraint{\footnote{At one loop order, the cubic interaction term will contribute to $\lambda$ by an amout $\Delta \lambda \sim g^2$. A renormalized value of $\lambda$ smaller than $g^2$ needs to be finely tuned at each order in perturbation theory, which is ``unnatural".}} 
$g^2 \, \sim \, \lambda$ and the value
$\lambda \sim 10^{-12}$ required if the CMB anisotropies induced by the model are to have their observed values, it follows that (for $\sigma < m_{pl}$)
\be
T_R \, < \, 10^{10} {\rm GeV} \, .
\ee
This would imply no GUT baryogenesis, no GUT-scale defect production, and no gravitino problems in supersymmetric models with $m_{3/2} > T_R$, where $m_{3/2}$ is the gravitino mass. As we shall see, these conclusions change radically if we adopt an improved analysis of reheating.

\smallskip
\noindent{\it Modern Theory of Reheating}

As was first realized in \cite{TB90}, the above analysis misses an essential point. To see this, we focus on the equation of motion for the matter field $\chi$ coupled to the inflaton $\varphi$ via the interaction Lagrangian ${\cal L}_I$ of (\ref{intlag}). Taking into account for the moment only the cubic interaction term, the equation of motion for the Fourier modes $\chi_k$ becomes
\be \label{reseq}
{\ddot \chi_k} + 3H{\dot \chi_k} + (k_p^2 + m_{\chi}^2 + 2g^2\sigma\varphi)\chi_k \, = \, 0 ,
\ee
where $k_p$ is the time-dependent physical wavenumber. 

Let us for the moment neglect the expansion of the Universe and assume the case of narrow resonance ($g^2 \sigma \varphi_0 < \omega^2$). In this case, the friction term in (\ref{reseq}) drops out and $k_p$ is time-independent, and Equation (\ref{reseq}) becomes a harmonic oscillator equation with a time-dependent mass determined by the dynamics of $\varphi$. In the reheating phase, $\varphi$ is undergoing oscillations. Thus, the mass in (\ref{reseq}) is varying periodically. In the mathematics literature, this equation is called the Mathieu equation. It is well known that there is an instability. In physics, the effect is known as {\bf parametric resonance} (see e.g. \cite{parres}). At frequencies $\omega_n$ corresponding to half integer multiples of the frequency $\omega$ of
the variation of the mass, i.e.
\be
\omega_k^2 = k_p^2 + m_{\chi}^2 \, = \, ({n \over 2} \omega)^2 \,\,\,\,\,\,\, n = 1, 2, ... ,
\ee
there are instability bands with widths $\Delta \omega_n$. For values of $\omega_k$ within the instability band, the value of $\chi_k$ increases exponentially:
\be
\chi_k \, \sim \, e^{\mu t} \,\,\,\, {\rm with} \,\,\, \mu \sim {{g^2 \sigma \varphi_0} \over {\omega}} \, ,
\ee
with $\varphi_0$ being the amplitude of the oscillation of $\varphi$. Since the widths of the instability bands decrease as a power of $g^2$ with increasing $n$, only the lowest instability band is important. Its width is
\be
\Delta \omega_k \, \sim \, g \sigma^{1/2} \varphi_0^{1/2} \, .
\ee
Note, in particular, that there is no ultraviolet divergence in computing the total energy transfer from the $\varphi$ to the $\chi$ field due to parametric resonance.

It is easy to include the effects of the expansion of the Universe (see e.g. \cite{TB90,KLS94,STB95}). The main effect is that the value of $\omega_k$ becomes time-dependent. Thus, a mode slowly enters and leaves the resonance bands. As a consequence, any mode lies in the resonance band for only a finite time. This implies that the calculation of energy transfer is perfectly well-behaved. No infinite time divergences arise.

It is now possible to estimate the rate of energy transfer, whose order of magnitude is given by the phase space volume of the lowest instability band multiplied by the rate of growth of the mode function $\chi_k$. Using as an initial condition for $\chi_k$ the value $\chi_k \sim H$ given by the magnitude of the expected quantum fluctuations, we obtain
\be \label{entransf}
{\dot \rho} \, \sim \, \mu ({\omega \over 2})^2 \Delta\omega_k H e^{\mu t} \, .
\ee

From (\ref{entransf}) it follows that provided that the condition
\be \label{rescond}
\mu \Delta t \, >> 1
\ee
is satisfied, where $\Delta t < H^{-1}$ is the time a mode spends in the instability band, then the energy transfer will procede fast on the time scale
of the expansion of the Universe. In this case, there will be explosive particle production, and the energy density in matter at the end of reheating will be given by the energy density at the end of inflation.  
 
Working out the time $\Delta t$ that a mode remains in the instability band for our model, expressing $H$ in terms of $\varphi_0$ and $m_{pl}$, and $\omega$ in terms of $\sigma$, and using the naturalness relation $g^2 \sim \lambda$, the condition for explosive particle production becomes
\be \label{rescond2}
{{\varphi_0 m_{pl}} \over {\sigma^2}} \, >> \, 1 \, ,
\ee
which is satisfied for all chaotic inflation models with $\sigma < m_{pl}$ (recall that slow rolling ends when $\varphi \sim m_{pl}$ and that therefore the initial amplitude $\varphi_0$ of oscillation is of the order $m_{pl}$).

We conclude that rather generically, reheating in chaotic inflation models will be explosive. This implies that the energy density after reheating will be approximately equal to the energy density at the end of the slow rolling period. Therefore, as suggested in \cite{KLS96,Tkachev} and \cite{KLR96}, respectively, GUT scale defects may be produced after reheating and GUT-scale baryogenesis scenarios may be realized, provided that the GUT energy scale is lower than
the energy scale at the end of slow rolling.

Note, however, that the state of $\chi$ after parametric resonance is {\bf not} a thermal state. The spectrum consists of high peaks in distinct wave bands. An important question which remains to be studied is how this state thermalizes.
For some interesting work on this issue see \cite{therm}. As emphasized in \cite{KLS96} and \cite{Tkachev}, the large peaks in the spectrum may lead to symmetry restoration and to the efficient production of topological defects (for a differing view on this issue see \cite{AC96,Boyan2}). Since the state after explosive particle production is not a thermal state, it is useful to follow
\cite{KLS94} and call this process ``preheating" instead of reheating.

A futher interesting conjecture which emerges from the parametric resonance analysis of preheating \cite{KLS94,STB95} is that the dark matter in the Universe may consist of remnant coherent oscillations of the inflaton field. In fact, it can easily be checked from (\ref{rescond2}) that the condition for efficient transfer of energy eventually breaks down when $\varphi_0$ has decreased to a sufficiently small value. For the model considered here, an order of magnitude calculation shows that the remnant oscillations may well contribute significantly to the present value of $\Omega$.

Note that the details of the analysis of preheating are quite model-dependent. In fact \cite{KLS94}, in many models one does not get the kind of ``narrow-band" resonance discussed here, but ``wide-band" resonance. In this case, the energy transfer is even more efficient.

There has recently been a lot of work on various aspects of reheating (see e.g. \cite{Yoshimura1,Boyan1,Kaiser,ALR96,Yoshimura2,ZMCB,Bassett} for different approaches). Many important questions, e.g. concerning thermalization and back-reaction effects during and after preheating (or parametric resonance) remain to be fully analyzed.

\section{Newtonian Theory of Cosmological Perturbations}

\subsection{Basic Concepts}

In inflationary Universe (and other particle physics-based) models, small amplitude seed perturbations are generated in the very early Universe.  They then grow by gravitational instability until the present time. In order
to be able to make the connection between particle physics and observations, it
is important to understand the gravitational evolution of fluctuations.  This
and the following section will summarize the theory of linear cosmological inhomogeneities. 
  
Both general relativity and quantum mechanics play a fundamental role.  
In inflationary Universe models, quantum effects seed  fluctuations, and thus a quantum analysis of the generation of fluctuations is essential. However, since the initial fluctuations are small, a linearized analysis is sufficient. Scales of cosmological interest today were larger than the Hubble radius for a long time interval during the early Universe. Hence, Newtonian gravity is inadequate to treat these perturbations, and general relativistic effects become essential.

In this section, after introducing some basic notation, we will summarize the main elements of the Newtonian theory of linear fluctuations. The next section will survey the full relativistic analysis.

Gravitational accretion is the mechanism by which small initial perturbations grow.  In a flat space-time background,  a density perturbation with
$\delta \rho > 0$ leads to an excess gravitational attractive
force acting on the surrounding matter.  This force is proportional to
$\delta \rho$, and will hence lead to exponential growth of the perturbation.
In an expanding background space-time, the acceleration is damped by the
expansion, which results in power-law increase of $\delta \rho$. The actual growth rate of inhomogeneities depends on the background cosmology (see e.g. \cite{Efstath,Padmanabhan} for modern reviews).

Because of our assumption that all perturbations start out with  a small
amplitude, we can linearize the equations for gravitational fluctuations.  The
analysis is then greatly simplified by going to momentum space in which all
modes $\delta (\underline{k})$ decouple.   

The ``power spectrum" $P(k)$ is defined by
\be \label{powersp}
P (k) = < |\delta (k) |^2 > \, , 
\ee
where the braces denote an ensemble average.
The physical measure of mass fluctuations on a length scale $\lambda$ is the
r.m.s. mass fluctuation $\delta M/M (\lambda)$ on this scale.  It is determined
by the power spectrum as follows:
\be \label{masspow}
< \big| {\delta M\over M} (\lambda) \big|^2 > \sim k_{\lambda}^3 P(k_{\lambda})
\, , 
\ee
where $k_{\lambda}$ is the wave number associated with $\lambda$ (see \cite{TASI,Korea} for explicit derivations of this relation).

If $P (k) \sim k^n$ then $n$ is called the index of the power spectrum.  For $n = 1$ we get the
so-called Harrison-Zel'dovich scale invariant spectrum \cite{HZ}.

Inflationary Universe (but also topological defect) models of structure formation predict a roughly scale invariant spectrum.  The distinguishing feature of this
spectrum is that the r.m.s. mass perturbations are independent of the scale $k$
when measured at the time $t_H (k)$ when the associated wavelength is equal to
the Hubble radius, i.e., when the scale ``enters" the Hubble radius.  Let us
derive this fact for the scales entering during the matter dominated epoch.
The time $t_H (k)$ is determined by
\be \label{Hubble}
k^{-1} a (t_H (k)) = t_H (k) 
\ee
which leads to $t_H (k) \sim k^{-3}$.
According to the linear theory of cosmological perturbations discussed in the
following subsection, the mass fluctuations increase as $a(t)$ for $t >
t_{eq}$.  Hence
\be \label{growth}
{\delta M\over M} (k, t_H (k)) = \left( {t_H (k)\over t} \right)^{2/3} \,
{\delta M\over M} \, (k, t) \sim {\rm const} \, , 
\ee
since the first factor scales (from (\ref{Hubble}) as $k^{-2}$ and -- using (\ref{masspow}) and inserting
$n=1$ -- the second as $k^2$.

\subsection{Newtonian Theory}

The Newtonian theory of cosmological perturbations is an approximate analysis
which is valid on wavelengths $\lambda$ much smaller than the Hubble radius $t$
and for negligible pressure $p$, i.e., $p \ll  \rho$.  It is based on expanding
the hydrodynamical equations about a homogeneous background solution.

The starting points are the continuity, Euler and Poisson equations
\be \label{Newteq1}
\dot \rho + \underline{\nabla} (\rho \underline{v}) = 0 
\ee
\be \label{Newteq2}
\underline{\dot v} + (\underline{v} \cdot \underline{\nabla}) \underline{v} = -
\underline{\nabla} \phi - {1\over \rho} \, \underline{\nabla} p 
\ee
\be  \label{Newteq3}
\nabla^2  \phi = 4 \pi G \rho 
\ee
for a fluid with energy density $\rho$, pressure $p$, velocity $\underline{v}$
and Newtonian gravitational potential $\phi$, written in terms of physical
coordinates $(t, \, \underline{r})$.

After some work (see \cite{Padmanabhan,Korea}), these equations can be combined to yield a second order differential equation for the fractional density contrast $\delta = \delta \rho / \rho$. Introducing comoving coordinates, expanding the above equations about a homogeneous background solution and assuming adiabatic perturbations, the resulting equation in Fourier space becomes
\be \label{momeq}
\ddot \delta_{\underline{k}} + 2 H \dot \delta_{\underline{k}} + \left( {c^2_s
k^2 \over a^2} - 4 \pi G \bar \rho \right) \delta_{\underline{k}} = 0 \, .
\ee
where $c_s$ is the speed of sound  
\be
c^2_s = {\partial p\over{\partial \rho}} \, ,  
\ee
$\bar \rho$ is the average energy density, and $\delta_{\underline{k}}$ stands for $\delta (\underline{k})$.

Already a quick look at Equation (\ref{momeq}) reveals the presence of a distinguished scale for cosmological perturbations, the Jeans length
$\lambda_J = 2 \pi / k_J$
with
\be \label{Jeans}
k^2_J = \left( {k\over a} \right)^2 = {4 \pi G \bar \rho\over c^2_s} \, .
\ee
On length scales larger than $\lambda_J$, the spatial gradient term is
negligible, and the term linear in $\delta$ in (\ref{momeq}) acts like a negative mass
square quadratic potential with damping due to the expansion of the Universe,
in agreement with the intuitive analysis at the begining of this section.  On length scales smaller than $\lambda_J$, however, (\ref{momeq}) becomes a damped harmonic oscillator equation and perturbations on these scales decay.

For $t > t_{eq}$ and for $\lambda \gg \lambda_J$, the spatial gradient term in (\ref{momeq}) can be neglected and in this approximation 
(\ref{momeq}) has the general solution
\be
\delta_k (t) = c_1 t^{2/3} + c_2 t^{-1} \, . 
\ee
This demonstrates that for $t > t_{eq}$ and $\lambda \gg \lambda_J$, the
dominant mode of perturbations increases as $a(t)$, a result we already used in
the previous subsection (see (\ref{growth})).

For $\lambda \ll \lambda_J$ and $t > t_{eq}$, the last term on the l.h.s. of   (\ref{momeq}) is negligible and the equation is that of a damped oscillator, i.e.
\be
\delta_k (t) \, \sim \, a^{-1/2} (t) \exp \{ \pm i c_s k \int dt^\prime a
(t^\prime)^{-1} \} \, . 
\ee

As an important application of the Newtonian theory of cosmological
perturbations, let us compare sub-horizon scale fluctuations in a
baryon-dominated Universe $(\Omega = \Omega_B = 1)$ and in a CDM-dominated
Universe with $\Omega_{CDM} = 0.9$ and $\Omega = 1$.  We consider scales which
enter the Hubble radius at about $t_{eq}$.

In the initial time interval $t_{eq} < t < t_{rec}$, the baryons are coupled to
the photons.  Hence, the baryonic fluid has a large pressure $p_B$
\be
p_B \simeq p_r = {1\over 3} \, \rho_r \, , 
\ee
and therefore the speed of sound is relativistic
\be
c_s \simeq \left( {p_r\over \rho_m} \right)^{1/2} = {1\over{\sqrt{3}}} \,
\left({\rho_r\over \rho_m} \right)^{1/2} \, . 
\ee
The value of $c_s$ slowly decreases in this time interval, attaining a value of
about $1/10$ at $t_{rec}$.  From (\ref{Jeans}) it follows that the Jeans mass $M_J$,
the mass inside a sphere of radius $\lambda_J$, increases until $t_{rec}$ when
it reaches its maximal value $M_J^{max}$
\be
M_J^{max} = M_J (t_{rec}) = {4 \pi\over 3} \, \lambda_J (t_{rec})^3 \bar \rho
(t_{rec}) \sim 10^{17} (\Omega h^2)^{-1/2} M_{\odot} \, . 
\ee

At the time of recombination, the baryons decouple from the radiation fluid.
Hence, the baryon pressure $p_B$ drops abruptly, as does the Jeans length (see
(\ref{Jeans})).  The remaining pressure $p_B$ is determined by the temperature and
thus continues to decrease as $t$ increases.  It can be shown that the Jeans
mass continues to decrease after $t_{rec}$, starting from a value
\be
M^-_J (t_{rec}) \sim 10^6 (\Omega h^2)^{-1/2} \, M_\odot 
\ee
(where the superscript ``$-$" indicates the mass immediately after $t_{eq}$.

In contrast, CDM has negligible pressure throughout the period $t > t_{eq}$ and
hence experiences no Jeans damping.  A CDM perturbation which enters the Hubble
radius at $t_{eq}$ with amplitude $\delta_i$ has an amplitude at $t_{rec}$
given by
\be
\delta^{CDM}_k (t_{rec}) \simeq \, {a (t_{rec})\over{a (t_{eq})}} \, \delta_i
\, , 
\ee
whereas a perturbation with the same scale and initial amplitude in a
baryon-dominated Universe is damped
\be
\delta_k^{BDM} (t_{rec}) \simeq \, \left({a (t_{rec})\over{a (t_{eq})}}
\right)^{-1/2} \, \delta_i \, . 
\ee
In order for the perturbations to have the same amplitude today, the initial
size of the inhomogeneity must be much larger in a BDM-dominated Universe than
in a CDM-dominated one:
\be
\delta^{BDM}_k (t_{eq}) \simeq \left( {z (t_{eq})\over{z (t_{rec})}}
\right)^{3/2} \delta_k^{CDM} \, (t_{eq}) \, . 
\ee
For $\Omega = 1$ and $h = 1/2$ the enhancement factor is about 30.

In a CDM-dominated Universe the baryons experience Jeans damping, but after
$t_{rec}$ the baryons quickly fall into the potential wells created by the CDM
perturbations, and hence the baryon perturbations are proportional to the CDM
inhomogeneities.  

The above considerations, coupled with information about CMB anisotropies, can
be used to rule out a model with $\Omega = \Omega_B = 1$.  The argument goes as
follows.  For adiabatic fluctuations, the amplitude of CMB
anisotropies on an angular scale $\vartheta$ is determined by the value of
$\delta \rho/\rho$ (strictly speaking, the relativistic potential $\Phi$ to be discussed in the following subsection) on the corresponding length scale $\lambda (\vartheta)$ at
$t_{eq}$:
\be
{\delta T\over T} (\vartheta)  = {1\over 3} \, {{\delta \rho} \over \rho} (
\lambda (\vartheta), \, t_{eq} ) \, . 
\ee
 On scales of clusters we know that (for $\Omega = 1$ and $h = 1/2$)
\be
\left({{\delta \rho} \over \rho} \right)_{CDM} \, (\lambda (\vartheta), \, t_{eq})
\simeq z (t_{eq})^{-1} \simeq 10^{-4} \, , 
\ee
using the fact that today on cluster scales $\delta \rho/\rho \simeq 1$.  The
bounds on $\delta T/ T$ on small angular scales are
\be
{\delta T\over T}(\vartheta)  << 10^{-4} \, , 
\ee
consistent with the predictions for a CDM model, but inconsistent with those of
a $\Omega = \Omega_B = 1$ model, according to which we would expect
anisotropies of the order of $10^{-3}$.  This is yet another argument in
support of the existence of nonbaryonic dark matter.

Two further important aspects of the Newtonian theory of cosmological perturbations which were covered in my lectures but which are omitted here due to lack of space (see e.g. \cite{Padmanabhan,Korea} for recent reviews) are the growth of matter fluctuations before $t_{eq}$ (they grow only logarithmically in time), and the suppression of hot dark matter inhomogeneities due to free streaming. 

\section{Relativistic Theory of Cosmological Perturbations}

\subsection{Classical Analysis}

On scales larger than the Hubble radius $(\lambda > t)$ the Newtonian theory of
cosmological perturbations obviously is inapplicable, and a general
relativistic analysis is needed.  On these scales, matter is essentially frozen
in comoving coordinates.  However, space-time fluctuations can still increase
in amplitude.

In principle, it is straightforward to work out the general relativistic theory
of linear fluctuations \cite{Lifshitz}.  We linearize the Einstein  equations
\be
G_{\mu\nu} = 8 \pi G T_{\mu\nu} 
\ee
(where $G_{\mu\nu}$ is the Einstein tensor associated with the space-time
metric $g_{\mu\nu}$, and $T_{\mu\nu}$ is the energy-momentum tensor of matter)
about an expanding FRW background $(g^{(0)}_{\mu\nu} ,\, \varphi^{(0)})$:
\begin{eqnarray}
g_{\mu\nu} (\underline{x}, t) & = & g^{(0)}_{\mu\nu} (t) + h_{\mu\nu}
(\underline{x}, t) \\
\varphi (\underline{x}, t) & = & \varphi^{(0)} (t) + \delta \varphi
(\underline{x}, t) \,  
\end{eqnarray}
and pick out the terms linear in $h_{\mu\nu}$ and $\delta \varphi$ to obtain
\be \label{linein}
\delta G_{\mu\nu} \> = \> 8 \pi G \delta T_{\mu\nu} \, . 
\ee
In the above, $h_{\mu\nu}$ is the perturbation in the metric and $\delta
\varphi$ is the fluctuation of the matter field $\varphi$.  We have denoted all
matter fields collectively by $\varphi$.

In practice, there are many complications which make this analysis highly
nontrivial.  The first problem is ``gauge invariance" \cite{PressVish}   Imagine starting
with a homogeneous FRW cosmology and introducing new coordinates which mix
$\underline{x}$ and $t$.  In terms of the new coordinates, the metric now looks
inhomogeneous.  The inhomogeneous piece of the metric, however, must be a pure
coordinate (or "gauge") artefact.  Thus, when analyzing relativistic
perturbations, care must be taken to factor out effects due to coordinate
transformations.

There are various methods of dealing with gauge artefacts.  The simplest and
most physical approach is to focus on gauge invariant variables, i.e.,
combinations of the metric and matter perturbations which are invariant under
linear coordinate transformations.

The gauge invariant theory of cosmological perturbations is in principle
straightforward, although technically rather tedious. In the following I will
summarize the main steps and refer the reader to \cite{MFB92} for the details and further references (see also \cite{MFBrev} for a pedagogical introduction and \cite{Bardeen,BKP83,KoSa84,Durrer,Lyth,Hwang,EllisBruni,Salopek} for other approaches).

We consider perturbations about a spatially flat Friedmann-Robertson-Walker
metric
\be
ds^2 = a^2 (\eta) (d\eta^2 - d \underline{x}^2) 
\ee
where $\eta$ is conformal time (related to cosmic time $t$ by $a(\eta)  d \eta
= dt$).  A scalar metric perturbation (see \cite{Stewart} for a precise definition)
can be written in terms of four free functions of space and time:
\be
\delta g_{\mu\nu} = a^2 (\eta) \pmatrix{2 \phi & -B_{,i} \cr
-B_{,i} & 2 (\psi \delta_{ij} + E_{,ij}) \cr} \, . 
\ee
Scalar metric perturbations are the only perturbations which couple to energy
density and pressure.

The next step is to consider infinitesimal coordinate transformations
which preserve the scalar nature of $\delta g_{\mu\nu}$, and to calculate the
induced transformations of $\phi, \psi, B$ and $E$.  Then we find invariant
combinations to linear order.  (Note that there are in general no combinations
which are invariant to all orders \cite{SteWa}.)  After some algebra, it follows
that
\begin{eqnarray}
\Phi & = & \phi + a^{-1} [(B - E^\prime) a]^\prime \\
\Psi & = & \psi - {a^\prime\over a} \, (B - E^\prime)  
\end{eqnarray}
are two invariant combinations (a prime denotes differentiation
with respect to $\eta$).

Perhaps the simplest way \cite{MFB92} to derive the equations of motion for gauge invariant variables is to consider the linearized
Einstein equations (\ref{linein}) and to write them out in the longitudinal gauge defined by $B = E = 0$, in which $\Phi = \phi$ and $\Psi = \psi$, to directly obtain gauge invariant equations.

For several types of matter, in particular for scalar field matter,  
$\delta T^i_j \sim \delta^i_j$ 
which implies $\Phi = \Psi$.  Hence, the scalar-type cosmological perturbations
can in this case be described by a single gauge invariant variable.  The
equation of motion takes the form \cite{BST,BK84,Lyth,TASI,Gotz} 
\be \label{conserv}
\dot \xi = O \left({k\over{aH}} \right)^2 H \xi  
\ee
where
\be
\xi = {2\over 3} \, {H^{-1} \dot \Phi + \Phi\over{1 + w}} + \Phi \, . 
\ee

The variable $w = p/ \rho$ (with $p$ and $\rho$ background pressure and energy
density respectively) is a measure of the background equation of state.  In
particular, on scales larger than the Hubble radius, the right hand side of
(\ref{conserv}) is negligible, and hence $\xi$ is constant.

If the equation of state of matter is constant, {\it i.e.}, $w = {\rm
const}$, then $\dot \xi = 0$ implies that the relativistic potential
is time-independent on scales larger than the Hubble radius, i.e. $\Phi (t) = {\rm const}$. During a transition from an
initial phase with $w = w_i$  to a phase with $w = w_f$, $\Phi$ changes. In many cases, a good approximation to the dynamics given by (\ref{conserv}) is
\be \label{cons3}
{\Phi\over{1 + w}}(t_i)  \, = \, {\Phi\over{1 + w}}(t_f)  \, , 
\ee

In order to make contact with matter perturbations and Newtonian intuition, it
is important to remark that,  as a consequence of the Einstein constraint
equations, at Hubble radius crossing $\Phi$ is a measure of the fractional
density fluctuations:
\be
\Phi (k, t_H (k) ) \sim {\delta \rho\over \rho} \, ( k , \, t_H (k) ) \, .
\ee
 
\subsection{Quantum Analysis}

As mentioned earlier, the primordial fluctuations in an inflationary cosmology are generated by quantum fluctuations. What follows is a very brief description of the unified
analysis of the quantum generation and evolution of perturbations in an inflationary Universe (for a detailed review see \cite{MFB92}).
The basic point is that at the linearized level, the equations describing both gravitational and matter perturbations can be quantized in a consistent way. The use of gauge invariant variables makes the analysis both physically clear and computationally simple. 

The first step of this analysis is to consider the action for the linear perturbations in a background homogeneous and isotropic Universe, and to express the result in terms of gauge invariant variables describing the fluctuations. Focusing on the scalar perturbations, it turns out that, after a lot of algebra, the action reduces to the action of a single gauge invariant free scalar field with a time dependent mass \cite{Mukh88,Sasaki}  (the time dependence reflects the expansion of the background space-time). This result is not surprising. Based on the study of classical cosmological perturbations, we know that there is only one field degree of freedom for the scalar perturbations. Since at the linearized level there are no mode interactions, the action for this field must be that of a free scalar field. 

The action thus has the same form as the action for a free scalar matter field in a time dependent gravitational or electromagnetic background, and we can use standard methods to quantize this theory (see e.g. \cite{Birrell}). If we employ canonical quantization, then the mode functions of the field operator obey the same classical equations as we derived in the gauge-invariant analysis of relativistic perturbations. 

The time dependence of the mass is reflected in the nontrivial form of the solutions of the mode equations. The mode equations have growing modes which correspond to particle production or equivalently to the generation and amplification of fluctuations. We can start the system off (e.g. at the beginning of inflation) in the vacuum state (defined as a state with no particles with respect to a local comoving observer). The state defined this way will not be the vacuum state from the point of view of an observer at a later time. The Bogoliubov mode mixing technique can be used to calculate the number density of particles at a later time. In particular, expectation values of field operators such as the power spectrum can be computed.

The resulting power spectrum gives the following result for the mass perturbations at time $t_i(k)$:
\be \label{inmass}
\left( {\delta M\over M} \right)^2 \, (k, t_i (k)) \sim k^3 \left({V^\prime
(\varphi_0) \delta \tilde \varphi (k, t_i (k))\over \rho_0} \right)^2 \sim
\left({V^\prime (\varphi_0) H\over \rho_0 } \right)^2 \, . 
\ee
 
If the background scalar field is rolling slowly, then
\be \label{slowroll1}
V^\prime (\varphi_0 (t_i (k))) =  3 H | \dot \varphi_0 (t_i (k)) | \, .
\ee
and
\be \label{slowroll2}
(1 + p/\rho)(t_i(k)) \, \simeq \, \rho_0^{-1} {\dot \varphi_0^2}(t_i(k)) \, .
\ee
Combining (\ref{cons3}), (\ref{inmass}), (\ref{slowroll1}) and (\ref{slowroll2}) and we get
\be
{\delta M\over M} (k, \, t_f (k))  \sim \, {3 H^2 | \dot \varphi_0
(t_i (k)) |\over{\dot \varphi^2_0 (t_i (k))}} =  {3H^2\over{| \dot \varphi_0 (t_i (k))|}} 
\ee
This result can now be evaluated for specific models of inflation to find the
conditions on the particle physics parameters which give a value
\be \label{obs2}
{\delta M\over M} (k, \, t_f (k))  \sim 10^{-5} 
\ee
which is required if quantum fluctuations from inflation are to provide the
seeds for galaxy formation and agree with the CMB anisotropy limits.

For chaotic inflation with a potential
\be
V (\varphi) = {1\over 2} m^2 \varphi^2 \, , 
\ee
we can solve the slow rolling equations for the inflaton to obtain
\be \label{massconstr}
{\delta M\over M} (k, t_f (k))  \sim 10 {m\over m_{pl}} 
\ee
which implies that $m \sim 10^{13} \, {\rm GeV}$ to agree with (\ref{obs2}).

Similarly, for a quartic potential  
\be
V (\varphi) = {1\over 4} \lambda \varphi^4 
\ee
we obtain
\be \label{lambdaconstr}
{\delta M\over M} (k, \, t_f (k)) \sim  10 \cdot \lambda^{1/2} 
\ee
which requires $\lambda \leq 10^{-12}$ in order not to conflict with observations.

The conditions (\ref{massconstr}) and (\ref{lambdaconstr}) require the presence of small parameters in the particle physics model.  
It  has been shown quite generally \cite{Freese}  that
small parameters are required if inflation is to solve the fluctuation problem.

To summarize the main results of the analysis of density
fluctuations in inflationary cosmology:
\begin{enumerate}
\item{} Quantum vacuum fluctuations in the de Sitter phase of an inflationary
Universe are the source of perturbations.
\item{} As a consequence of the change in the background equation of state, the evolution outside the Hubble radius produces a large
amplification of the perturbations.  In fact, unless the particle physics
model contains very small coupling constants, the predicted fluctuations are
in excess of those allowed by the bounds on cosmic microwave anisotropies.
\item{} The quantum generation and classical evolution of fluctuations can be treated in a unified manner. The formalism is no more complicated that the study of a free scalar field in a time dependent background.
\item{} Inflationary Universe models generically produce an approximately scale invariant Harrison-Zel'dovich spectrum
\be
{\delta M\over M} (k , t_f (k) ) \, \simeq \, {\rm const.} 
\ee
\end{enumerate}

It is not hard to construct models which give a different spectrum.  All that
is required is a significant change in $H$ during the period of
inflation. 

\section{Problems of Inflation}

In spite of its great success at resolving some of the problems of standard cosmology and at providing a causal, predictive theory of structure formation, there are several important unresolved conceptual problems in inflationary cosmology. In the following, I will single out four major problems and 
mention a couple of new ideas which may help solve some of these issues.

\subsection{A Partial List of Conceptual Problems}

{\it Fluctuation Problem}: A generic problem for all realizations of inflation studied up to now concerns the amplitude of the density perturbations which are induced by quantum fluctuations during the period of exponential expansion \cite{flucts,BST}. From the amplitude of CMB anisotropies measured by COBE, and from the present amplitude of density inhomogeneities on scales of clusters of galaxies, it follows that the amplitude of the mass fluctuations ${\delta M} / M$ on a length scale given by the comoving wavenumber $k$ at the time $t_H(k)$ when that scale crosses the Hubble radius in the FRW period is of the order $10^{-5}$. 
 However, as was discussed in detail in the previous section, the present realizations of inflation based on scalar quantum field matter generically \cite{Freese} predict a much larger value of these fluctuations, unless a parameter in the scalar field potential takes on a very small value. There have been many attempts to justify such small parameters based on specific particle physics models, but no single convincing model has emerged.

{\it Super-Planck Scale Problem}: In many models of chaotic inflation, the period of inflation is so long that comoving scales of cosmological interest today corresponded to a physical wave number way in excess of the Planck scale at the beginning of inflation. In extrapolating the evolution of cosmological perturbations according to linear theory to these very early times, we are implicitly making the assumption that the theory remains perturbative to arbitrarily high energies. If there were completely new physics at the Planck scale, the predictions might change. A similar concern about black hole Hawking radiation has been raised in \cite{Jacobson}.

{\it Cosmological Constant Problem}:
Since the cosmological constant acts as an effective energy density, its value is bounded from above by the present energy density of the Universe. In Planck units, the constraint on the effective cosmological constant $\Lambda_{eff}$ is
(see e.g. \cite{cosmorev})
\be
{{\Lambda_{eff}} \over {m_{pl}^4}} \, \le \, 10^{- 122} \, .
\ee
This constraint applies both to the bare cosmological constant and to any matter contribution which acts as an effective cosmological constant.

The true vacuum value of the potential $V(\varphi)$ acts as an effective cosmological constant. Its value is not constrained by any particle physics requirements (in the absence of special symmetries). The cosmological constant problem is thus even more accute in inflationary cosmology than it usually is. The same unknown mechanism which must act to shift the potential such that inflation occurs in the false vacuum must also adjust the potential to vanish in the true vacuum. 
Supersymmetric theories may provide a resolution of this problem, since unbroken supersymmetry forces $V(\varphi) = 0$ in the supersymmetric vacuum. However, supersymmetry breaking will induce a nonvanishing $V(\varphi)$ in the true vacuum after supersymmetry breaking.

{\it Singularity Problem}: Scalar field-driven inflation does not eliminate singularities from cosmology. Although the standard assumptions of the Penrose-Hawking theorems break down if matter has an equation of state with negative pressure, as is the case during inflation, nevertheless it can be shown that an initial singularity persists in inflationary cosmology \cite{Borde}.
One way to interpret this result is that the theory is still incomplete and lacking an important ingredient.

A possible way to find solutions to some of the above problems is to look for realizations of inflation which do not make use of scalar fields. One such alternate way to obtain inflation \cite{ARZ} is by replacing the fundamental scalar field by a condensate (see \cite{Ball} and \cite{Parker} for different approaches to this problem). An advantage in following this route to inflation is that the symmetry breaking mechanisms observed in nature (in condensed matter systems) are induced by the formation of condensates such as Cooper pairs. In a condensate model the energy function is determined by the confining dynamics. The situation with respect to the cosmological constant problem is therefore improved. However, the bare (classical-level) cosmological constant problem still persists.
The main problem of studying the possibility of obtaining inflation using condensates is that the quantum effects which determine the theory are highly nonperturbative. In particular, the effective potential written in terms of a condensate  does not correspond to a renormalizable theory and will in general \cite{ARZ}  contain terms of arbitrary power of the condensate. However (see \cite{BZ96}), one may make progress by assuming certain general properties of the effective potential. In \cite{ARZ} it was demonstrated that it is in principle possible to obtain inflation from condensates.  

Another way to obtain inflation is by making use of higher derivative gravity theories. In fact, the first
model with exponential expansion of the Universe was obtained \cite{Starob}  in an $R^2$ gravity theory. The extra degrees of freedom associated with the higher derivative terms act as scalar fields with a potential which automatically vanishes in the true vacuum. In this context, it is also possible to consider the possibility of obtaining a nonsingular cosmology.

\subsection{Inflation and Nonsingular Cosmology}

The question we wish to address in this subsection is whether it is 
possible to construct a class of effective actions for gravity which 
have improved singularity properties and which predict inflation, 
with the constraint that they give the correct low curvature limit.
Since Planck scale physics will generate corrections to the Einstein action, it is quite reasonable to consider higher derivative gravity models.

What follows is a summary of recent work \cite{MB92}  in which  
an effective action for gravity in which all solutions 
with sufficient symmetry are nonsingular was proposed.  The theory is a higher 
derivative modification of the Einstein action, and is obtained by 
a constructive procedure well motivated in analogy with the analysis 
of point particle motion in special relativity.  The resulting theory 
is asymptotically free in a sense which will be specified below.  

The goal is to construct a theory with the property that the metric 
$g_{\mu\nu}$ approaches the de Sitter metric $g_{\mu\nu}^{DS}$, a 
metric with maximal symmetry which admits a geodesically complete and 
nonsingular extension, as the curvature $R$ approaches the Planck 
value $R_{pl}$.  Here, $R$ stands for any curvature invariant.  
Naturally, from classical considerations, $R_{pl}$ is a free 
parameter.  However, if the theory is connected with Planck scale 
physics, $R_{pl}$ will be given by the Planck (or string) scale.

The construction has some very appealing 
consequences.  Consider, for example, a collapsing spatially 
homogeneous Universe.  According to Einstein's theory, this Universe 
will collapse in finite proper time to a final ``big crunch" singularity. 
In the new theory, however, the Universe will approach a de Sitter model as 
the curvature increases.  If the 
Universe is closed, there will be a de Sitter bounce followed by 
re-expansion.  Similarly, spherically 
symmetric vacuum  solutions would be nonsingular, i.e., black holes 
would have no singularities in their centers. Outside and even slightly inside the horizon, the structure of a 
large black hole would be unchanged compared to what is predicted by 
Einstein's theory, since 
all curvature invariants are small in those regions.  However, for $r \rightarrow 0$ 
(where $r$ is the radial Schwarzschild coordinate), the solution 
changes and approaches a de Sitter solution.  This 
would have interesting consequences for the black hole information 
loss problem.

To motivate the effective action for a nonsingular cosmology, we turn to an  analogy, point particle motion in the theory of special relativity.
The transition from the Newtonian theory of point particle motion to 
the special relativistic theory transforms a theory with no bound on 
the velocity into one in which there is a limiting velocity, the speed 
of light $c$ (in the following we use units in which $\hbar = c = 1$).  
This transition can be obtained \cite{MB92}  by starting with the action of a point particle with world line $x(t)$:
\be
S_{\rm old} = \int dt {1\over 2} \dot x^2 \, , 
\ee
introducing \cite{Alt}  a Lagrange multiplier field $\varphi$ which couples to $\dot x^2$, the quantity to be made finite, and which has a potential 
$V(\varphi)$. The new action is
\be
S_{\rm new} = \int dt \left[ {1\over 2} \dot x^2 + \varphi \dot x^2 - 
V (\varphi) \right] \, . 
\ee
From the constraint equation
\be
\dot x^2 = {\partial V\over{\partial \varphi}} \, , 
\ee
it follows that $\dot x^2$ is limited provided $V(\varphi)$ increases 
no faster than linearly in $\varphi$ for large $|\varphi|$.  The small 
$\varphi$ asymptotics of $V(\varphi)$ is determined by demanding that 
at low velocities the correct Newtonian limit results:
\begin{eqnarray} 
V (\varphi) \, \sim \, \varphi^2 \>\,\,\, & {\rm as} & \> |\varphi| 
\rightarrow 0 \, , \label{ascond1} \\
V (\varphi) \sim \varphi \>\,\,\, & {\rm as}  & \> |\varphi| \rightarrow \infty 
\, . \label{ascond2} 
\end{eqnarray}
Choosing the simple interpolating potential
\be
V (\varphi) \, = \, {2 \varphi^2\over{1 + 2 \varphi}} \, , 
\ee
the Lagrange multiplier can be integrated out, resulting in the well-known
action
\be
S_{\rm new} \, = \, {1\over 2} \int dt \sqrt{1 - \dot x^2} 
\ee
for point particle motion in special relativity.

The procedure for obtaining a nonsingular Universe theory \cite{MB92}  is based on generalizing the above Lagrange multiplier construction to gravity.  
Starting from the Einstein action, we can introduce a Lagrange 
multiplier $\varphi$ coupled to a curvature invariant $I_2(g_{\mu \nu})$ 
which has the property that it only vanishes for de Sitter metrics. The potential for $\varphi$ is chosen such that for small $|\varphi|$ the equations of general relativity are recovered, but such that for large $|\varphi|$, $I_2$ is driven to zero (i.e. the metric becomes de Sitter). The new action is
\be
S \, = \, \int d^4 x \sqrt{-g} (R + \varphi \, I_2 + V (\varphi) ) \, . 
\ee
For metrics with special symmetries (including cosmological and Schwarschild space-times) it is possible to find an invariant $I_2$ which satisfies the criteria. $I_2$ is a combination of quadratic curvature invariants.
For a detailed analysis of this model, the reader is referred to \cite{MB92,BMS93}. In these papers it is shown that 
all homogeneous and isotropic solutions are nonsingular. 
It can also be shown that the two-dimensional Schwarzschild black hole is nonsingular \cite{BMT93}.
 
By construction, all solutions are de Sitter at high curvature.  Thus, 
the theories automatically have a period of inflation (driven by the 
gravity sector in analogy to Starobinsky inflation \cite{Starob}) in the 
early Universe.
A very important property of these theories is asymptotic freedom.  This 
means that the coupling between matter and gravity goes to zero at 
high curvature, and might lead to an automatic suppression mechanism 
for scalar fluctuations.

\subsection{Inflation and Gravitational Back-Reaction}
 
It has been discovered \cite{ABM2}  that in an inflationary Universe, the back-reaction of long wavelength cosmological fluctuations acts as a negative cosmological constant. This effect may lead to a dynamical relaxation of the effective cosmological constant, as was also proposed by Tsamis and Woodard \cite{Woodard} (who considered infrared quantum effects of gravitational waves).

Gravitational back-reaction of cosmological perturbations concerns itself with the evolution of space-times which consist of small fluctuations about a symmetric Friedmann-Robertson-Walker space-time with metric $g_{\mu \nu}^{(0)}$. The goal is to study the evolution of spatial averages of observables in the perturbed space-time. In linear theory, such averaged quantities evolve like the corresponding variables in the background space-time. However, beyond linear theory perturbations have an effect on the averaged quantities. In the case of gravitational waves, this effect is well-known \cite{Brill}: gravitational waves carry energy and momentum which effects the background in which they propagate. Here, we shall focus on scalar metric perturbations.

The analysis of gravitational back-reaction \cite{ABM1} is related to early work by Brill, Hartle and Isaacson \cite{Brill}, among others. The idea is to expand the Einstein equations to second order in the perturbations, to assume that the first order terms satisfy the equations of motion for linearized cosmological perturbations \cite{MFB92}  (hence these terms cancel), to take the spatial average of the remaining terms, and to regard the resulting equations as equations for a new homogeneous metric $g_{\mu \nu}^{(0, br)}$ which includes the effect of the perturbations to quadratic order:
\be \label{breq}
G_{\mu \nu}(g_{\alpha \beta}^{(0, br)}) \, = \, 8 \pi G \left[ T_{\mu \nu}^{(0)} + \tau_{\mu \nu} \right]\,
\ee
where the effective energy-momentum tensor $\tau_{\mu \nu}$ of gravitational back-reaction contains the terms resulting from spatial averaging of the second order metric and matter perturbations:
\be  \label{efftmunu}
\tau_{\mu \nu} \, = \, < T_{\mu \nu}^{(2)} - {1 \over {8 \pi G}} G_{\mu \nu}^{(2)} > \, ,
\ee
where pointed brackets stand for spatial averaging, and the superscripts indicate the order in perturbations.

As formulated in (\ref{breq}) and (\ref{efftmunu}), the back-reaction problem is not independent of the coordinatization of space-time and hence is not well defined. It is possible to take a homogeneous and isotropic space-time, choose different coordinates, and obtain a nonvanishing $\tau_{\mu \nu}$. This ``gauge" problem is related to the fact that in the above prescription, the hypersurface over which the average is taken depends on the choice of coordinates. 

The key to resolving the gauge problem is to realize that to second order in perturbations, the background variables change. A gauge independent form of the back-reaction equation (\ref{breq}) can hence be derived \cite{ABM1} by defining background and perturbation variables $Q = Q^{(0)} + \delta Q$ which do not change under linear coordinate transformations. Here, $Q$ represents collectively both metric and matter variables. The gauge-invariant form of the back-reaction equation then looks formally identical to (\ref{breq}), except that all variables are replaced by the corresponding gauge-invariant ones. We will follow the notation of \cite{MFB92}, and use as gauge-invariant perturbation variables the Bardeen potentials \cite{Bardeen} $\Phi$ and $\Psi$ which in longitudinal gauge coincide with the actual metric perturbations $\delta g_{\mu \nu}$. Calculations hence simplify greatly if we work directly in  longitudinal gauge.
 
In \cite{ABM2}, the effective energy-momentum tensor $\tau_{\mu \nu}$ of gravitational back-reaction was evaluated for long wavelength fluctuations in an inflationary Universe in which the matter responsible for inflation is a scalar field $\varphi$ with the potential
\be
V(\varphi) \, = \, {1 \over 2} m^2 \varphi^2 \, .
\ee
Since in this model there is no anisotropic stress, the perturbed metric in longitudinal gauge can be written \cite{MFB92}  in terms of a
single gravitational potential $\phi$
\be
ds^2 =  (1+ 2 \phi) dt^2 - a(t)^2(1 - 2\phi) \delta_{i j} dx^i dx^j  \, ,
\ee
where $a(t)$ is the cosmological scale factor. 

It is now straightforward to compute $G_{\mu \nu}^{(2)}$ and 
$T_ {\mu \nu}^{(2)}$ in terms of the background fields and the metric and matter fluctuations $\phi$ and $\delta \varphi$, By taking averages and making use of (\ref{efftmunu}), the effective energy-momentum tensor $\tau_{\mu \nu}$ can be computed \cite{ABM2}.

The general expressions for the effective energy density $\rho^{(2)} = \tau^0_0$ and effective pressure $p^{(2)} = - {1 \over 3} \tau^i_i$ involve many terms. However, they greatly simplify if we consider perturbations with wavelength greater than the Hubble radius. In this case, all terms involving spatial gradients are negligible. From the theory of linear cosmological perturbations (see e.g. \cite{MFBrev}) it follows that on scales larger than the Hubble radius the time derivative of $\phi$ is also negligible as long as the equation of state of the background does not change. The Einstein constraint equations relate the two perturbation variables $\phi$ and $\delta \varphi$, enabling scalar metric and matter fluctuations to be described in terms of a single gauge-invariant potential $\phi$. During the slow-rolling period of the inflationary Universe, the constraint equation takes on a very simple form and implies that $\phi$ and $\delta \varphi$ are proportional. The upshot of these considerations is that $\tau_{\mu \nu}$ is proportional to the two point function $< \phi^2 >$, with a coefficient tensor which depends on the background dynamics. In the slow-rolling approximation we obtain \cite{ABM2} 
\be
\rho^{(2)} \, \simeq \, - 4 V < \phi^2 >
\ee
and
\be
p^{(2)} \, = \, - \rho^{(2)} \, .
\ee
This demonstrates that the effective energy-momentum tensor of long-wavelength cosmological perturbations has the same form as a negative cosmological constant.

\section*{Acknowledgments}

I wish to thank the organizers for putting together an excellent school, and for inviting me to speak. I also thank my collaborators, in particular Raul Abramo and Slava Mukhanov, for collaboration on some of the work reported on at the end of these lecture notes. This work has been supported in part by the U.S. Department of Energy under Contract DE-FG0291ER40688, Task A; and by an NSF 
collaborative research award NSF-INT-9312335.

\end{document}